\documentclass[reprint,superscriptaddress,amsmath,amssymb,aps,prb]{revtex4-2}

\usepackage[version=3]{mhchem}
\usepackage{subfigure}
\usepackage{graphicx}\usepackage{outlines}\usepackage{dcolumn}\usepackage{bm}\usepackage{hyperref}\usepackage{braket}
\usepackage{soul}

\usepackage{xcolor}

\newcommand{\sub}[1]{\ensuremath{_{\mathrm{#1}}}}

\begin{document}

\title{Magnetic-field dependence of spin-phonon relaxation and dephasing\\
due to $g$-factor fluctuations from first principles}

\author{Joshua Quinton}
\affiliation{Department of Physics, Applied Physics, and Astronomy, Rensselaer Polytechnic Institute, Troy, New York 12180, USA}

\author{Mayada Fadel}
\affiliation{Department of Materials Science and Engineering, Rensselaer Polytechnic Institute, Troy, New York 12180, USA}

\author{Junqing Xu}
\affiliation{Department of Physics, Hefei University of Technology, 420 Feicui Road, University City, Hefei Economic and Technological Development Zone, Anhui Province, China}

\author{Adela Habib}
\affiliation{Theoretical Division, Los Alamos National Laboratory, Los Alamos, NM 87545, USA}

\author{Mani Chandra}
\affiliation{Department of Materials Science and Engineering, Rensselaer Polytechnic Institute, Troy, New York 12180, USA}

\author{Yuan Ping}
\email{yping3@wisc.edu}
\affiliation{Department of Materials Science and Engineering, University of Wisconsin, Madison, WI 53706, USA}

\author{Ravishankar Sundararaman}
\email{sundar@rpi.edu}
\affiliation{Department of Materials Science and Engineering, Rensselaer Polytechnic Institute, Troy, New York 12180, USA}

\date{\today}

\begin{abstract}
The electron spin decay lifetime in materials can be characterized by relaxation (T$_1$) and
irreversible (T$_2$) and reversible decoherence processes (T$_2$*).
Their interplay leads to a complex dependence of spin lifetimes on the direction and magnitude of magnetic fields, relevant for spintronics and quantum information applications.
Here, we use real-time first-principles density matrix dynamics simulations to directly simulate Hahn echo measurements, disentangle dephasing from decoherence, and predict $T_1$, $T_2$ and $T_2^\ast$ spin lifetimes.
We show that $g$-factor fluctuations lead to non-trivial magnetic field dependence of each of these lifetimes in inversion-symmetric crystals of CsPbBr$_3$ and silicon, even when only intrinsic spin-phonon scattering is present.
Most importantly, fluctuations in the off-diagonal components of the $g$-tensor lead to a strong magnetic field dependence of even the $T_1$ lifetime in silicon.
Our calculations elucidate the detailed role of anisotropic $g$-factors in determining the spin dynamics even in simple, low spin-orbit coupling materials such as silicon.
\end{abstract}

\maketitle \section{Introduction} 

The manipulation and control of long-lived electron spins in materials is vital for spintronic and quantum information technologies \cite{RevModPhys.76.323, Awschalom2018, HIROHATA2020166711}, necessitating a detailed understanding of the relaxation mechanisms limiting spin lifetimes.
Experimentally, ultrafast pump-probe spectroscopy techniques map non-equilibrium spin dynamics with excellent time resolution, and have been used to quantify spin relaxation in high spin-orbit coupling materials suitable for spin control \cite{Chen2021, Giovanni2015}.
The complexity of spin dynamics in materials, however, requires theoretical modeling to disentangle the interplay of several competing mechanisms.

Theoretical models including the Elliot-Yafet and the D'yakonov-Perel models successfully capture limiting cases of spin relaxation dynamics in inversion-symmetric and inversion-symmetry-broken crystals with well-known spin textures such as Rashba and Dresselhaus \cite{PhysRev.96.266, DP1972}. Previous works have developed models for $T_1$, $T_2$, and $T_2^*$ lifetimes for specific materials in quantum well and quantum dot heterostructures, however these are material-specific\cite{PhysRevB.70.195314, LU2007501, PhysRevB.87.161305}.
The general case of complex spin textures in promising bulk materials with long-lived spins, such as halide perovskites, requires first-principles modeling to capture the transition between these limiting mechanisms and elucidate the complete dependence of spin relaxation on temperature and magnetic field.
We have recently demonstrated first-principles density matrix dynamics, combining coherent evolution with Lindbladian electron-phonon, electron-electron and electron-defect scattering, to simulate spin dynamics in a variety of materials spanning simple semiconductors (Si, GaAs), 2D materials (graphene, transition metal chalcogenides) and halide perovskites \cite{Xu2020, Xu2021,
doi:10.1021/acs.nanolett.1c03345, PhysRevB.105.115122, Xu2023, Xu2024}.

First-principles spin dynamics simulations allow isolating the contributions of different scattering mechanisms to the $T_1$ time, in the absence of magnetic fields or with magnetic fields parallel to the spin direction, as well as to the $T_2^\ast$ time of spins precessing in a perpendicular magnetic field \cite{Xu2023}.
However, in this latter case, $T_2^\ast$ includes reversible dephasing contributions due to difference in precession frequencies of individual spins, in addition to the irreversible decoherence of each spin, characterized by the $T_2$ time.
The $T_2$ decoherence time characterizes the decay time of a single spin prepared in a coherent superposition, and is the relevant quantity for quantum information applications, while measurements and simulations on ensemble of spins typically include dephasing and therefore characterize $T_2^\ast$.

In this Article, we directly simulate Hahn spin-echo measurements \cite{PhysRev.80.580, Balasubramanian2009} using real-time first-principles density matrix simulations of spin dynamics to separate dephasing from decoherence and predict $T_2$ in addition to $T_1$ and $T_2^\ast$ of spins in materials.
Specifically, we simulate the magnetic-field dependent intrinsic spin-phonon relaxation in silicon and CsPbBr$_3$, varying vastly in unit cell complexity and spin-orbit coupling strength.
Halide perovskites such as CsPbBr$_3$ with strong spin-orbit coupling are attractive for spintronics \cite{Ping2018}, and their $T_1$ and $T_2^\ast$ have been recently simulated from first principles \cite{Xu2024}, but the contributions of dephasing and decoherence ($T_2$) are yet to be investigated.
We show that variations in the electron Lande $g$-factor lead to dephasing and significant differences among $T_1$, $T_2^\ast$ and $T_2$ in these materials. 
Our results highlight the impact of the anisotropic $g$-tensor in spin dynamics, leading to an unexpectedly strong magnetic field dependence of even the $T_1$ lifetime in silicon.

\section{Theory and Computational Methods}

\subsection{First-principles density matrix dynamics}

We perform real-time simulations of first-principles density matrix dynamics,
combining coherent Liouville evolution of the electronic density matrix $\rho$
with Lindbladian electron-phonon scattering given by
\begin{multline}
\frac{\partial\rho_{\alpha_{1}\alpha_{2}}}{\partial t}
= -\frac{i}{\hbar}\left[H', \rho\right]_{\alpha_{1}\alpha_{2}} + \frac{2\pi}{\hbar N_{q}}
\sum_{q\lambda\pm\alpha'\alpha'_{1}\alpha'_{2}}n_{q\lambda}^{\pm} \\
\times \mathrm{Re} \left[\begin{array}{c}
\left(I-\rho\right)_{\alpha_{1}\alpha'} A^{q\lambda\pm}_{\alpha'\alpha'_{1}} \rho_{\alpha'_{1}\alpha'_{2}} A^{q\lambda\mp}_{\alpha'_{2}\alpha_{2}}\\
-A^{q\lambda\mp}_{\alpha_{1}\alpha'} \left(I-\rho\right)_{\alpha'\alpha'_{1}} A^{q\lambda\pm}_{\alpha'_{1}\alpha'_{2}} \rho_{\alpha'_{2}\alpha_{2}}
\end{array}\right].\label{eq:Lindblad}
\end{multline}
where $\alpha \equiv (\mathbf{k},n)$ specifies the electron wavevector
$\mathbf{k}$ and band $n$, $\pm$ denotes absorption and emission of phonons with
wave vector $q = \mp(k - k')$ and mode index $\lambda$,  $n_{q\lambda}^\pm \equiv n_{q\lambda} +
\frac{1}{2} \pm \frac{1}{2}$, and $N_{q} = N_{k}$ is the total
number of electron and phonon wave vectors sampled in the Brillouin zone.
Finally, $A^{q\lambda \pm}_{\alpha \alpha'} = g_{\alpha
\alpha'}^{q\lambda\pm} \delta_G ^{1/2}(\epsilon_\alpha - \epsilon_{\alpha'} \pm
\hbar \omega_{q \lambda}) \exp(i(\epsilon_\alpha - \epsilon_{\alpha'})t)$, where
$g^{q\lambda\pm}_{\alpha \alpha'}$ is the electron-phonon matrix element,
$\epsilon_\alpha$ are the electron energies, and the $\delta_G$-function (regulated by a Gaussian function)
enforces
energy conservation. 
Note that the density matrix $\rho$ is in the interaction picture, analytically accounting for the time evolution due to the electronic energies $\epsilon_\alpha$, and the explicit coherent evolution in the first term above is only due to an additional perturbing Hamiltonian $H'$, such as due to interactions with external fields.
The above has been obtained by tracing over the phonon
degrees of freedom in the quantum Liouville equation and then applying the
Born-Markov approximation in a form of Lindbladian dynamics for electronic system-phonon bath interactions\cite{Taj2009}.

The electron and phonon energies and scattering matrix elements are obtained from density-functional theory (DFT) using
the JDFTx code \cite{SUNDARARAMAN2017278}, using the Perdew-Burke-Ernzerhof exchange-correlation functional \cite{PhysRevLett.77.3865} with fully-relativistic normconserving pseudopotentials  \cite{PhysRevB.88.085117}.
Since such self-consistent spin-orbit coupling calculations are spinorial, the spin degrees of freedom are implicitly captured within the band indices within Eq.~\ref{eq:Lindblad}.
The DFT calculations are performed on coarse electronic-$\mathbf{k}$ and phonon-$\mathbf{q}$ meshes, and then interpolated onto much finer meshes using maximally localized Wannier functions.
From the solution $\rho(t)$ of Eq.~\ref{eq:Lindblad}, we can calculate time-dependent expectation values of observables such as spin, starting from an initial $\rho_0$ in equilibrium at a specified electron chemical potential and test magnetic field to introduce a free carrier density with a net initial paramagnetic spin polarization.
Specifically, to capture electron spin dynamics below, we set the electron chemical potential just below the conduction band minimum for each material.
This first-principles framework naturally captures both Elliot-Yafet (EY) relaxation due to spin flips and Dyakonov-Perel (DP) spin relaxation in inversion-symmetry broken systems on an equal footing \cite{DP1972, PhysRev.96.266, yafet1963g}.
See \cite{Xu2021} for further details on the computational framework and benchmarks of this approach, and section I of the SI \cite{SI} for further details on the DFT calculations.

\subsection{Interaction with magnetic fields}

The Zeeman perturbation Hamiltonian due to an external magnetic field $\mathbf{B}$ is 
\begin{equation}
H'_{\mathbf{k}nn'} = \mu_B\mathbf{B}\cdot(\mathbf{L}_{\mathbf{k}nn'} + |g_0| \mathbf{S}_{\mathbf{k}nn'}),
\label{equ:hamiltonian}
\end{equation}
where $\mathbf{L}_{\mathbf{k}nn'}$ and $\mathbf{S}_{\mathbf{k}nn'}$ are the orbital angular momentum and spin matrix elements calculated from DFT, $\mu_B$ is Bohr magneton and $g_0 \approx -2.002$ is the free-electron g-factor.
In the real-time simulations below, we introduce both constant background magnetic fields and time-dependent fields for the Hahn echo setup.
For numerical stability, we include the effect of the time-independent magnetic fields within the reference Hamiltonian used for defining the interaction picture, and only treat the time-dependent part of $H'$ using the first term in Eq.~\ref{eq:Lindblad}.
(This does not change the results, but allows using a larger time step in the simulations.)

The computation of the matrix elements of the orbital angular momentum $\mathbf{L}$ in periodic systems is non-trivial.
We use the techniques detailed in \cite{Multunas2023} to calculate them both directly from DFT and using Wannier interpolation.
Briefly, we use $k$-derivatives of the DFT Bloch functions $u_{\mathbf{k}n}(\mathbf{r})$ to calculate $\mathbf{L}_{\mathbf{k}nn'}
= (\langle i\nabla_{\mathbf{k}}u_{\mathbf{k}n}|
\times \hat{\mathbf{p}} |u_{\mathbf{k}n'}\rangle + h.c.)/2$,
where $\hat{\mathbf{p}}$ is the momentum operator.
This fully accounts for long-range components of the angular momentum in solids (and does not involve any additional approximations such as restricting the contributions to atom neighborhoods).
In general, we need Wannier interpolation in the first-principles density matrix dynamics method for fine sampling of the Brillouin zone, but sharp features in the reciprocal space variation of $\mathbf{L}$ leads to large Wannier interpolation errors.
To address this issue, we calculate energies and matrix elements of momentum $\mathbf{P}$ and spin $\mathbf{S}$ using Wannier interpolation in a first pass while identifying all the $\mathbf{k}$ with states close to the band edges / Fermi level that contribute to the dynamics.
We then replace all electronic energies and matrix elements for the selected $\mathbf{k}$ with explicit DFT calculations, while only the electron-phonon matrix elements depend on the Wannier interpolation.
In doing this replacement, we identify the undetermined phases and unitary rotations within degenerate subspaces between the DFT and Wannier, as detailed in Section II of the SI \cite{SI}.

\begin{figure*}
\includegraphics[width=\linewidth]{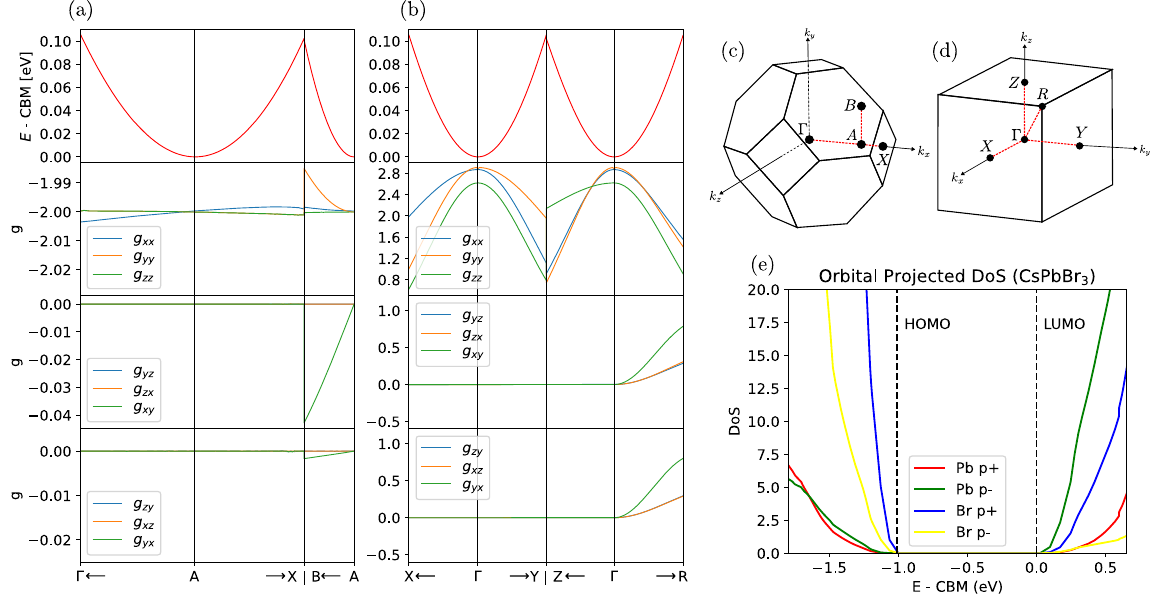}
\caption{$g$-tensor components for (a) Si and (b) CsPbBr$_3$ along $k$-paths close to the conduction band minimum, indicated in the corresponding Brillouin zones (c) and (d) respectively.
The top panels of (a,b) show the electron energies, while the remaining panels show the diagonal and off-diagonal components of the $g$-tensor.
Si exhibits $g$ close to the free-electron $g_0$ due to the weak spin-orbit coupling, while CsPbBr$_3$ shows strong variations in $g$ with even the sign changing relative to $g_0$ due to the Pb $p-$ states dominating the conduction states, as shown in (e) the orbital-projected density of states in CsPbBr$_3$.}
\label{fig:g_matrix}
\end{figure*}

\subsection{Lande g-factors}

The dynamics with the Zeeman perturbation Hamiltonian described above automatically accounts for the change of the Lande $g$-factor with electronic state due to the non-trivial orbital angular momentum of the electrons.
However, it is useful to extract and visualize the state-dependent $g$-factors to interpret the resulting spin dynamics.
In crystals, the $g$-factor is generally tensorial \cite{PhysRevLett.85.369} and strongly dependent on the electronic structure \cite{Sharma2024}.  
Fluctuations in $g$-factor with elctronic state affect the Larmor precession frequency of spins in the presence of a transverse magnetic field, leading to the dephasing of spins in spin relaxation \cite{Xu2024}.
Here, we show that the tensorial nature of the $g$-factor leads to complex magnetic field effects not just for transverse magnetic fields ($T_2^\ast$ and $T_2$), but also in the longitudinal $T_1$ spin relaxation that nominally should not involve any spin precession effects.

To extract the $g$-tensor from $\mathbf{S}$ and $\mathbf{L}$ matrix elements, we need to correctly account for the arbitrary degenerate-subspace unitary rotations in the DFT calculations.
We first construct Pauli matrices in the frame of the spin matrix elements for a pair of spin bands as
\begin{equation}
\Tilde{\sigma}^k_i = S^i_{\mathbf{k}} / \sqrt{\frac{1}{2} \mathrm{Tr}(S^i_{\mathbf{k}} S^i_{\mathbf{k}})}
\end{equation}
where $S^i_{\mathbf{k}}$ is the $2\times 2$ spin matrix elements for that pair of bands in Cartesian direction $i$.
We can then define the coefficients of the $g$-tensor by expanding  $-L^i_{\mathbf{k}} + g_0 S^i_{\mathbf{k}} = \sum_j g^{ij}_{\mathbf{k}} \Tilde{\sigma}^j_{\mathbf{k}}/2$, which can be inverted using the trace relations of the Pauli matrices to
\begin{equation}
\label{g-tensor2}
g^{ij}_{\mathbf{k}} = \mathrm{Tr}((-L^i_{\mathbf{k}} + g_0 S^i_{\mathbf{k}}) \cdot \Tilde{\sigma}^j_{\mathbf{k}}).
\end{equation}
Note that this $3\times 3$ $g$-tensor for each electronic state captures the interaction of a magnetic field in the $i$ direction with a spin in the $j$ direction.
Also, note that the negative sign on $L$ above is due to the negative charge on the electron, which is why $g_0$ is also negative in the standard sign convention.

Figure~\ref{fig:g_matrix} shows the resulting matrix elements of $g$ for states within 0.1~eV of the conduction band minimum in the two prototype materials considered in this work, silicon and CsPbBr$_3$.
The $g$ factor for Si remains very close to the free-electron $g_0 \approx -2.002$, as expected due to its low spin-orbit coupling.
Importantly, note that the off-diagonal components of $g$ along the lower-symmetry segments of the $k$-path are of a comparable and greater magnitude than the variation of the diagonal components from $g_0$, suggesting that these off-diagonal components may play a greater role in the field-dependent spin dynamics in Si.

In contrast, the $g$ factors overall vary much more strongly with $k$ in CsPbBr$_3$ due to the high spin-orbit coupling, with slightly stronger variations in the diagonal components.
Most importantly, note that the sign of $g$ is opposite to $g_0$, which is expected due to the dominant Pb $p-$ orbital character of the conduction band edge states as seen in the orbital-projected density of states (Figure~\ref{fig:g_matrix}(e)); the net angular momentum of the $p-$ states with $j=1/2$ and $l=1$ is dominated by $L$ and is in the opposite direction to $S$.
Additionally, both the sign and magnitude of the electron $g$ factors are in excellent agreement with experimental measurements for lead halide perovskites \cite{gHybridPerovskites}.
Based on these results, we expect strong dephasing due to large variations in the Larmor precession frequency in CsPbBr$_3$, and due to the effects dominated by the change in the precession axis due to off-diagonal $g$ matrix elements in Si.

\begin{figure*}
\begin{center}
\includegraphics[width=\linewidth]{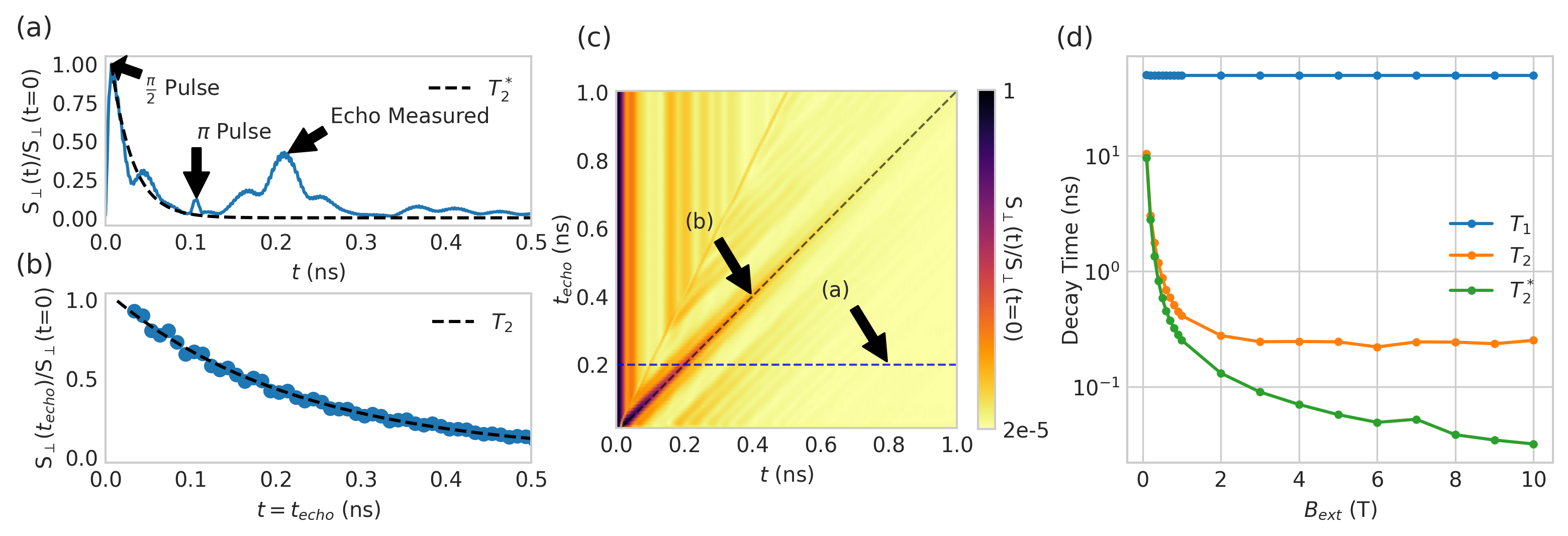}
\caption{
Predictions of spin relaxation for CsPbBr$_3$ at 4~K, and a magnetic field $B = 10$~T for (a-c).
(a) Spin magnitude in the plane of precession during a Hahn echo simulation, showing the spin initialization at $t=0$ using a $\pi/2$ pulse of magnetic field rotating in the precession plane, spin flip at 0.1~ns with a $\pi$ pulse and the corresponding echo at 0.2~ns.
(b) Echo magnitude as a function of echo time, obtained from a sequence of simulations as in (a), eliminates the dephasing and allows extraction of $T_2$ from an exponential fit.
(c) Spin magnitude for several Hahn echo simulations collected together in horizontal slices, shows the $T_2^\ast$ relaxation time scale horizontally and the $T_2$ relaxation time scale along the echos on the diagonal. 
(d) All spin relaxation times $T_1$, $T_2$ and $T_2^\ast$ predicted as a function of magnetic field strength, indicating $T_1 \gg T_2 > T_2^\ast$ for the simulated field strengths, even when only accounting for electron-phonon spin relaxation with $g$-factor fluctuations.}
\label{fig:cpb_t2_colormap}
\end{center}
\end{figure*}

\subsection{Extraction of $T_1$, $T_2^\ast$ and $T_2$}

To measure spin lifetimes, we initialize a spin-polarized density matrix at $t=0$, simulate the real-time dynamics using Eq.~\ref{eq:Lindblad} and calculate the spin expectation values as a function of time.
For each case, the initial spin polarization is created by equilibrium in an external magnetic field, with $\rho(t=0) = f_{\mu,T}(H_0 + H')$, where $f_{\mu,T}$ is the Fermi function, $H_0$ is the reference electronic Hamiltonian and $H'$ is the interaction with external field $\mathbf{B}$ set to $\mathbf{B}_\mathrm{init}$.
Note that $\mathbf{B}_\mathrm{init}$ is only used to initialize the spin polarization and is not present for the dynamics at $t > 0$.
When we apply the constant magnetic field $\mathbf{B}$ in the dynamics after $t=0$ either zero or parallel to $\mathbf{B}_\mathrm{init}$, we expect the spin polarization to relax exponentially without any precession since the field and spin are always parallel to each other and obtain $T_1$ from fitting $\langle S(t)\rangle = S_0 \exp(-t/T_1)$.

On the other hand, if the constant field is perpendicular to the initial spin polarization, the spins precess in the field in addition to relaxing due to electron-phonon scattering.
For definiteness, consider the case when the spins are initialized with $\mathbf{B}_\mathrm{init} \parallel \hat{x}$ and the subsequent $\mathbf{B} \parallel \hat{z}$.
The spins then precess in the $xy$-plane and the net spin magnitude in the plane perpendicular to the field, $S_\perp(t) \equiv \sqrt{\langle S_x(t)\rangle^2 + \langle S_y(t)\rangle^2}$, decays both due to dephasing from differences in precession frequency and overall spin relaxation.
We therefore obtain $T_2^\ast$ by fitting $S_\perp(t) = S_0 \exp(-t/T_2^\ast)$.

Finally, to extract $T_2$, we directly simulate a Hahn spin echo measurement.
For the case with $\mathbf{B} \parallel \hat{z}$, we initialize spins in equilibrium with $\mathbf{B}_\mathrm{init} \parallel \hat{z}$ as well.
We now apply an additional time-dependent, rotating magnetic field $\mathbf{B}_1(t) = B_1 (\hat{x}\cos\Omega t + \hat{y}\sin\Omega t)$ for short pulses that serve to rotate the spins from $\hat{z}$ to $xy$-plane where spins precess with average Larmor frequency $\Omega = e|g|B/2m_e$.
Specifically, when the rotating field is on, the spins rotate away from the $z$-axis with angular frequency $\Omega_1 = e|g|B_1/2m_e$ and therefore, turning on $\mathbf{B}_1(t)$ for a duration of $\pi/(2\Omega_1)$ -- a `$\pi/2$' pulse -- rotates the spins into the $xy$-plane.

Figure~\ref{fig:cpb_t2_colormap}(a) shows the magnitude of the $xy$-plane spin as a function of time for CsPbBr$_3$, starting with the spins rotating into the plane due to the $\pi/2$ pulse.
The spin magnitude then decays over time due to both dephasing and decoherence, until a second $\pi$ pulse is applied after a delay time $\tau$.
This $\pi$ pulse rotates the directions of the spins by 180$^\circ$, thereby reversing them.
Hence, after the $\pi$ pulse, the reversible dephasing portion of the dynamics gets reversed, leading to a refocussing of the spins after another delay of $\tau$ and the spin `echo' after delay $t\sub{echo} = 2\tau$ since the first pulse.
However, the irreversible decoherence processes have occured for time $t\sub{echo}$ and hence the magnitude of the echo is smaller than the initial spin magnitude.
Consequently, plotting the echo magnitude versus $t\sub{echo}$ obtained from simulations with several different $\tau$ delays (Figure~\ref{fig:cpb_t2_colormap}(b)) isolates only the irreversible decoherence portion, and we can extract $T_2$ by fitting $S\sub{echo}(t\sub{echo}) = S_0 \exp(-t\sub{echo}/T_2)$.

Figure~\ref{fig:cpb_t2_colormap}(c) shows the results of several different echo time simulations altogether in a 2D map of spin versus time $t$ on the $x$-axis and echo $t\sub{echo}$ on the $y$-axis.
In this form, each horizontal slice is a singe simulation like the one shown in Figure~\ref{fig:cpb_t2_colormap}(a).
The spin echos show up at $t=t\sub{echo}$, so that the diagonal slice indicated corresponds to decay of the spin echo magnitude, \textrm{i.e.}, to Figure~\ref{fig:cpb_t2_colormap}(b).
The times at which the $\pi$ pulse are applied appear as a line with two times the slope of the echo signal.
The faint vertical stripes that appear above this line are from a non-monotonic dephasing signal caused by the finite $k$ resolution, limited by the computational cost of the first-principles density matrix dynamics simulations; these lines would disappear with increasing $k$ resolution.
Similarly, the faint lines parallel to the echo signal are reflections of this behavior after the spin trajectories are reversed by the $\pi$ pulse.
Essentially, in this plot, the time scale of decay in the horizontal direction corresponds to $T_2^\ast$, including both reversible dephasing and irreversible decoherence processes, while that along the diagonal direction only contains the irreversible decoherence processes and corresponds to $T_2$.
We repeat this entire sequence of $T_1$, $T_2$ and $T_2^\ast$ calculations for several different magnetic field magnitudes $B$ for each material, as we discuss below in Section~\ref{sec:Results}.

\section{Results}\label{sec:Results}

\subsection{CsPbBr$_3$} \label{cpb}

We start with predictions of all three spin relaxation times in CsPbBr$_3$ at 4 K temperature, extracting $T_1$ and $T_2^\ast$ from real-time simulations with field parallel and perpendicular to the spin, and $T_2$ from simulations of Hahn echo, as dicsussed above.
CsPbBr$_3$ has spatial inversion symmetry and time-reversal symmetry (in the absence of fields), and should therefore be governed by the Elliott-Yafet mechanism of spin relaxation.
Correspondingly, we expect spin-phonon relaxation to yield no field dependence in $T_1$, but a reduction of $T_2^\ast$ and $T_2$ spin lifetimes with field strength due to $g$-factor variations.

Figure \ref{fig:cpb_t2_colormap}(d) compares the predicted $T_1$, $T_2$ and $T_2^\ast$ as a function of applied magnetic field strength.
At zero field, all three lifetimes are expected to be similar, but at finite field strengths, $T_1$ reduces only marginally, whereas $T_2^\ast$ and $T_2$ reduce dramatically as expected.
The predicted values of $T_1$ and $T_2^\ast$ agree well with previous predictions and measurements \cite{Xu2024}.
The $k$-dependent fluctuation in orbital angular momentum -- and therefore effective $g$-factor -- leads to electron spins at different $k$ precessing at different rates in the external magnetic field, leading to the rapid reduction in $T_2^\ast$ with field strength, scaling roughly as $1/B$.
However, once we remove the dephasing effects and consider $T_2$, the spin lifetime decreases substantially compared to $T_1$ due to decoherence, but saturates at a finite value for large field strengths.
Consequently, this leads to a regime of $T_1 \gg T_2 > T_2^\ast$ for relevant field strengths in the range of a few Teslas.

\subsection{Tensorial $g$-factor effects on spin relaxation}

\begin{figure}
\begin{center}
\includegraphics[width=0.45\columnwidth]{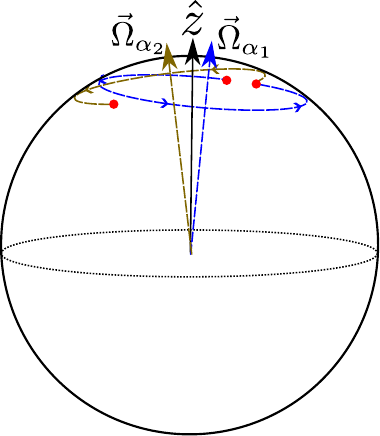}
\includegraphics[width=0.45\columnwidth]{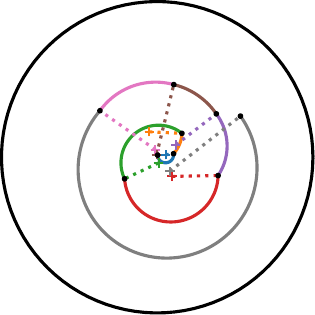}
\end{center}
\caption{
Impact of tensorial $g$-factor on spin-phonon relaxation.
(a) When the magnetic field is applied along the $z$ direction,
each spin $\hat{\mathbf{S}}_\alpha$ (where $\alpha$ indexes $k$ and band)
shown on the Bloch sphere precesses about a slightly different
Larmor precession axis $\hat{\mathbf{\Omega}}_\alpha$ with
corresponding frequency $\bar\Omega_{\alpha}=g_{\alpha}\mu_B B/\hbar$
due to the off-diagonal and diagonal elements of $g_{\alpha}$ respectively.
Upon each spin-phonon scattering, both the axis and frequency of precession change.
(b) The net trajectory of the spin seen in a top view of the Bloch sphere
exhibits a random walk away from $\hat{z}$, even when the spin is parallel to the field initially,
leading to a magnetic field dependence of even $T_1$ due to off-diagonal $g$ components,
in addition to those of $T_2$ and $T_2^\ast$ that are more commonly expected.}
\label{fig:bloch_sphere_model}
\end{figure}

To understand the magnetic field dependence of the spin lifetimes,
let us analyze the trajectories of spins with varying $g$-tensors
on the Bloch sphere (Figure~\ref{fig:bloch_sphere_model}).
When the field is applied along, say, the $z$ direction,
the Larmor precession frequency $\vec{\Omega} = g_{\alpha}\cdot\mu_B \vec{B}/\hbar$ will vary both in the axis direction and magnitude due to the variation of the off-diagonal and diagonal elements of $g_{\alpha}$ with $\alpha = \vec{k}$ and band.
Let $\sigma_\parallel$ and $\sigma_\perp$ be the standard deviations of the diagonal and off-diagonal components of $g_\alpha$ over all the band-edge electronic states, weighted by their thermal occupations.
We would then have a variance $\langle\delta\Omega\rangle^2 = (\sigma_\parallel\mu_B B/\hbar)^2$
in the precession frequency of the electron spins, and a similar variance proportional to
$\sigma_\perp^2$ in the axis of precession.

During collision events occuring on the timescale of momentum scattering $\tau_p$, 
the electron scatters to a new state $\alpha$, with a new $g_\alpha$.
With these assumptions, the spin adopts a circular arc on the Bloch sphere between scattering events,
and changes direction at each scattering event due to change of the precession axis,
leading to a random walk on the Bloch sphere over long time scales
as illustrated in Figure~\ref{fig:bloch_sphere_model}.
We can analytically compute these circular arcs, compute the average position of the spin on the Bloch sphere after many scattering events due to this random walk, and therefore predict all three spin relaxation times, as detailed in section III of the SI \cite{SI}.
In summary, we can show that
\begin{align}
T_1^{-1}
&= \tau_{s0}^{-1}
+ \frac{2\sigma_\perp^2}{\bar{g}^2\tau_p} F_\perp(\bar{g}\tau_p\mu_B B/\hbar)
\label{equ:t1_model}\\
T_2^{-1}
&= \tau_{s0}^{-1}
+ \frac{\sigma_\perp^2}{\bar{g}^2\tau_p} F_\perp(\bar{g}\tau_p\mu_B B/\hbar)
+ \frac{F_\parallel(\sigma_\parallel\tau_p\mu_B B/\hbar)}{\tau_p}
\label{equ:t2_model}\\
(T_2^\ast)^{-1}
&= T_2^{-1}
+ \frac{c_0 \sigma_\parallel \mu_B B}{\hbar}
    F_\parallel(\sigma_\parallel\tau_p\mu_B B/\hbar)
\label{equ:t2s_model}
\end{align}
where $\tau_{s0}$ is the zero-field Elliott-Yafet spin relaxation time due
to spin flips, $\bar{g}$ is the average diagonal $g$ factor, and $c_0$ is a
numerical constant for dephasing derived assuming a parabolic model for
$g$-factor variation near the band edge.
Above, the dimensionless field dependence functions
\begin{align}
F_\perp(x) &\equiv \frac{x^2}{1 + x^2} \quad\mathrm{and} \label{equ:Fperp}\\
F_\parallel(x) &\equiv 1 - \frac{\sqrt{\pi}e^{1/(2x^2)}}{x\sqrt{2}} \mathrm{erfc}\frac{1}{x\sqrt{2}}
\end{align}
corresponding to the contributions from off-diagonal ($\sigma_\perp$) and diagonal ($\sigma_\parallel$) fluctuations of the $g$-tensor respectively, are both $\approx x^2$ for $x \ll 1$ and $\to 1$ for $x \to \infty$.

\begin{figure}
\includegraphics[width=\columnwidth]{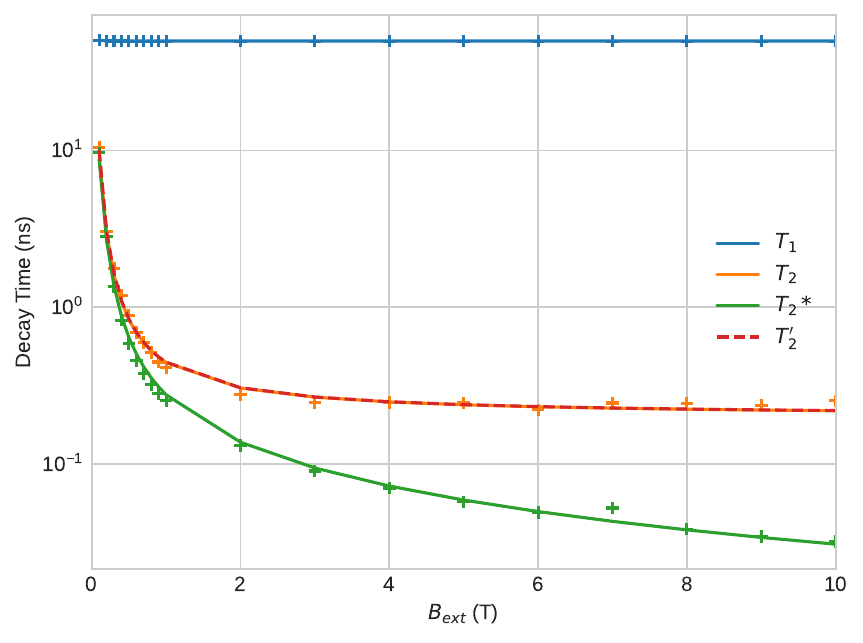}
\caption{
First-principles predictions of spin relaxation times in CsPbBr$_3$ as a functon of field strength,
compared to the analytical model accounting for spin trajectories due to $g$-tensor variations
(\ref{equ:t1_model} - \ref{equ:t2s_model}).
The DFT-derived and fit parameters of the analytical model are compared in Table~\ref{tab:cpb_model_params}.
Note that only $T_2$ and $T_2^\ast$ show substantial variation with field indicating the importance of
diagonal $g$-factor fluctuations ($\sigma_\parallel$ in the analytical model), rather than off-diagonal components.
$T'_2$, which removes $T_1$ contributions from $T_2$, almost coincides with $T_2$ here, because $T_1 \gg T_2$ throughout the relevant field range.
\label{fig:cpb_model_fits}}
\end{figure}

\begin{table}
\begin{tabular}{ |p{2cm}|p{2cm}|p{2cm}|  }
\hline
\textbf{Parameter} & \textbf{DFT Value} & \textbf{Fit Value} \\
\hline
$\tau_{s0}$ & 65.0 ns & - \\
\hline
$\bar{g}$ & 2.547 & - \\
\hline
$\tau_{p}$ & 0.222 ns & 0.200 ns \\
\hline
$\sigma_{\perp}$ & 0.0571 & 0.0551 \\
\hline
$\sigma_{\parallel}$ & 0.101 & 0.0788 \\
\hline
\end{tabular}
\caption{Comparison between DFT-predicted and fit parameters of the analytical model given by equations (\ref{equ:t1_model} - \ref{equ:t2s_model}) for CsPbBr$_3$.}
\label{tab:cpb_model_params}
\end{table}

Note that $T_2$ includes the effects contributing to $T_1$, and since $T_2 \le 2T_1$, it is useful to examine $(T_2')^{-1} \equiv T_2^{-1} - (2T_1)^{-1}$ where the $T_1$ contributions are explicitly subtracted out.
The analytical model yields
\begin{equation}
(T_2')^{-1}
= (2\tau_{s0})^{-1}
+ \frac{F_\parallel(\sigma_\parallel\tau_p\mu_B B/\hbar)}{\tau_p},
\label{equ:t2p_model}
\end{equation}
showing that the off-diagonal ($\sigma_\perp$) $g$-tensor contributions drop out from $T_2'$, leaving behind only the Elliot-Yafet spin flip rate and the diagonal ($\sigma_\parallel$) $g$-tensor contributions.

The above equations are fully general, recovering the correct limit for a wide range of scattering time $\tau_p$ and magnetic fields.
In the limit of zero field, all three times are equal $T_1 = T_2 = T_2^\ast = \tau_{s0}$ (and $T_2' = 2\tau_{s0}$).
For weak fields or strong scattering, with $\tau_p\Omega \equiv |\bar{g}|\tau_p\mu_BB/\hbar \ll 1$, all three inverse times are dominated by a $\tau_p B^2$ term, with prefactor proportional to $\sigma_\perp^2$, as expected for the Dyakonov Perel regime, but this is due to the change of precession axis.
Additionally, $T_2^{-1}$ and $(T_2^\ast)^{-1}$ exhibit another $\tau_p B^2$ term from Eq.~\ref{equ:t2_model} with prefactor proportional to $\sigma_\parallel^2$, which is the usual Dyakonov-Perel regime due to fluctuation of precession frequency, which occurs for  $\tau_p\Delta\Omega \equiv \sigma_\parallel\tau_p\mu_BB/\hbar \ll 1$, consistent with the past studies~\cite{Xu2024,WU201061}.

With increasing magnetic field $B$, $T_1$ eventually saturates to a constant given by
\begin{equation}
T_1^{-1} \rightarrow \tau_{s 0}^{-1} + \frac{2\sigma^2_\perp}{\bar{g}^2\tau_p}
\quad\mathrm{for}\quad
B \gg \frac{\hbar}{|\bar{g}|\tau_p\mu_B},
\label{equ:T1limit}
\end{equation}
while $T_2$ saturates at even larger fields to a constant given by
\begin{equation}
T_2^{-1} \rightarrow \tau_{s 0}^{-1}
+  \left(\frac{\sigma^2_\perp}{\bar{g}^2} + 1\right) \frac{1}{\tau_p}
\quad\mathrm{for}\quad
B \gg \frac{\hbar}{\sigma_\parallel\tau_p\mu_B}.
\label{equ:T2limit}
\end{equation}
Finally, in this large B (or weak scattering) limit, from Eq.~\ref{equ:t2s_model}, $(T^*_2)^{-1} \to c_0\sigma_\parallel\mu_BB/\hbar$, depending linearly on B and independent of $\tau_p$, as expected for the free induction decay regime.
Note that all of these field dependences are \emph{intrinsic}, arising only from $g$-tensor fluctuations and electron-phonon scattering; defects and nuclear-spin scattering would contribute additionally to the field dependence of spin relaxation in general.

Figure~\ref{fig:cpb_model_fits} compares the predictions of this model to the first-principles 
spin relaxation times for CsPbBr$_3$, with the corresponding paramters of the model compared
to the DFT predictions in Table~\ref{tab:cpb_model_params}.
Note that the model captures all the features of the field dependence of $T_1$, $T_2$ and $T_2^\ast$, with parameters quite close to the values extracted from DFT, confirming the proposed mechanism for the field dependence.
Note that $T_1$ does not change much with field strength with the high field limit reached for $B \gg 0.02$~T from Eq.~\ref{equ:T1limit}.
In contrast, $T_2$ saturates for $B \gg 0.6$~T from Eq.~\ref{equ:T2limit}, which is the transition clearly seen for magnetic fields on the 1 - 10~T scale in Fig.~\ref{fig:cpb_model_fits}.
At these fields, $T_2^\ast$ enters its free induction decay regime, and hence we see all three times $T_1$, $T_2$ and $T_2^\ast$ being distinct for CsPbBr$_3$.
Finally, since $T_1 \gg T_2$ over the entire field range, $T_2' \approx T_2$, which from the analytical model (Eq.~\ref{equ:t2s_model}) emphasizes that the $\sigma_\perp$ contributions are less important for CsPbBr$_3$ at this field range than those of $\sigma_\parallel$.

\subsection{Silicon}

\begin{figure}
\includegraphics[width=\columnwidth]{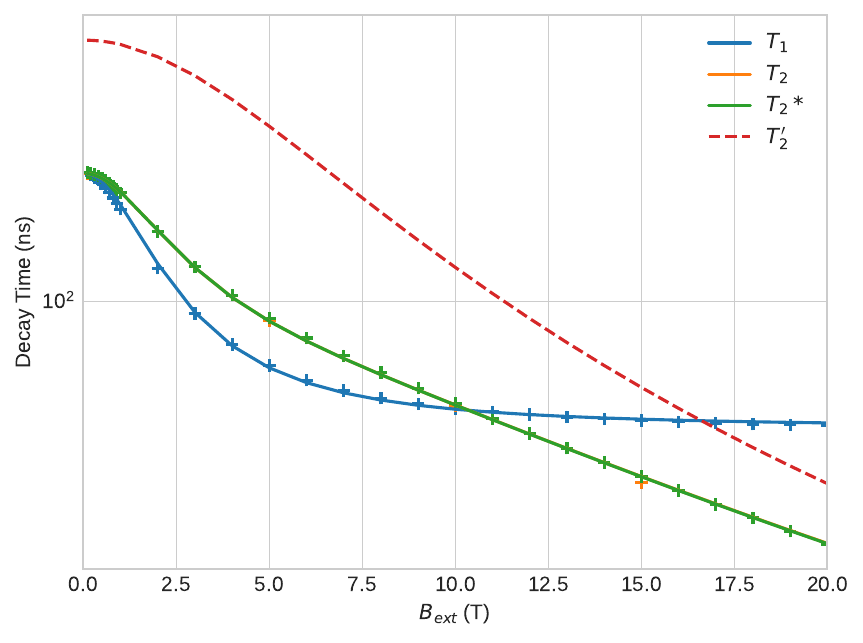}
\caption{Same as Figure~\ref{fig:cpb_model_fits}, but for silicon.
In this case, $T_1$ varies strongly with field as well, indicating the importance
of off-diagonal $g$-factor components ($\sigma_\perp$ in the analytical model)
in addition to the diagonal ones.
Note that $T_2 \approx T_2^\ast$ due to the small $\sigma_\parallel$ resulting
in very weak dephasing at relevant field strengths.
In contrast, $T_2' > T_2$ since $T_1$ effects are significant within $T_2$ in this field range for Si.
The DFT-derived and fit parameters of the analytical model are compared
in Table~\ref{tab:si_model_params}.}
\label{fig:si_t1_t2s}
\end{figure}

Unlike CsPbBr$_3$, silicon is a material with weak g-factor fluctuation and therefore minimal dephasing.
Based on \textit{ab initio} results, we observe significant magnetic field dependence of both the $T_1$ and $T_2^\ast$ spin relaxation in silicon (Figure~\ref{fig:si_t1_t2s}), shown for a temperature of 100~K, selected such that the zero-field spin relaxation times are at the same order of magnitude as the CsPbBr$_3$ case.

\begin{table}
\begin{tabular}{ |p{2cm}|p{2cm}|p{2cm}|  }
\hline
\textbf{Parameter} & \textbf{DFT Value} & \textbf{Fit Value} \\
\hline
$\tau_{s0}$ & 194.77 ns & - \\
\hline
$\bar{g}$ & 2.000 & - \\
\hline
$\tau_{p}$ & 0.00207 ns & 0.00150 ns \\
\hline
$\sigma_{\perp}$ & 0.00614 & 0.00652 \\
\hline
$\sigma_{\parallel}$ & 0.00119 & 0.00224 \\
\hline
\end{tabular}
\label{tab:si_model_params}
\caption{Comparison between DFT-predicted and fit parameters of the analytical model
given by equations (\ref{equ:t1_model} - \ref{equ:t2s_model}) for silicon.}
\end{table}

Based on the analytical model, we see that the unexpected field dependence of $T_1$ 
is due the off-diagonal elements of the $g$-tensor, captured by $\sigma_\perp$ in the model.
The net $T_1$ is due to two parallel mechanisms: an EY spin-flip relaxation with rate $\tau_{s0}^{-1}$ and a DP-like random walk on the Bloch sphere due to the off-diagonal $g$-tensor elements discussed above.
The EY spin flip rate $\tau_{s0}^{-1}$ is independent of magnetic field strength,
and is much greater than the momentum scattering rate ($\tau_p$).
At zero field, the EY spin relaxation dominates, however the DP-like
mechanism introduces a magnetic field dependence in $T_1$ and $T_2$ due to the off-diagonal components of the $g$-tensor.

$T_1$ saturates to a constant for $B \gg 2.7$~T from Eq.~\ref{equ:T1limit}, which is visible on the relevant field scales of 1 - 20~T considered in Fig.~\ref{fig:si_t1_t2s} (unlike the CsPbBr$_3$ case where the transition occurred at much smaller fields).
In contrast, $T_2$ will saturate only for $B \gg 4600$~T, far beyond the field scales considered in Fig.~\ref{fig:si_t1_t2s}.
We therefore also see $T_2^\ast \approx T_2$ in Fig.~\ref{fig:si_t1_t2s}, because we are substantially below the field scales of the free induction decay regime.
Instead, we now see $T_2'$ deviating significantly from $T_2$ because the off-diagonal $g$-tensor ($\sigma_\perp$) effects, which contributs to $T_2$ but are removed from $T_2'$, are significant for Si in this field range.
Consequently, due to this substantial change in field scales, we see qualitatively different behavior for Si compared to CsPbBr$_3$, even though they are both well-described by the same analytic model.

Finally, comparing the zero-field limit of CsPbBr$_3$ and silicon (Figures \ref{fig:cpb_model_fits} and \ref{fig:si_t1_t2s} respectively), both $T_2$ and $T_2^\ast$ approach the zero-field $T_1$ -- at this limit these times all become equivalent to the EY spin flip time.
Since CsPbBr$_3$ is a material with strong SOC, it is expected that the magnetic field dependence of $T_2$ and $T_2^\ast$ is significantly stronger even in the $10^{-1}$ T regime compared to silicon.
This is also supported by equations \ref{equ:t2_model} and \ref{equ:t2s_model} and the parameters calculated from first principles.

\section{Conclusions}

Spin coherence time $T_2$, an experimentally challenging timescale to measure, can be reliably
calculated \textit{ab initio} using real-time simulations of a spin-echo
procedure. This allows reversible dephasing effects to be separated from the
irreversible relaxation mechanisms which contribute to the $T_2^\ast$ lifetime.
This can be used to analyze detailed magnetic field interactions of spin
decoherence in materials using parameters calculated from first-principles.

Our results show that small off-diagonal components of the $g$-tensor -- leading to
small variance in the Larmor precession axis of spins in the presence of an
external magnetic field -- can lead to a significant effect in transverse spin
relaxation and also, surprisingly, longitudinal spin lifetime $T_1$. This
leads to a mechanism for external magnetic field dependence of $T_1$,
which can emerge even in materials with inversion-symmetry and negligible
internal magnetic fields. By analyzing the off-diagonal elements of the $g$-factor
as calculated from first principles, the precise magnetic field dependence of
spin relaxation times can be estimated with reasonable accuracy. 

This work highlights the complex interplay of electron-phonon relaxation
and anisotropic $g$-tensor effects from the band structure on the intrinsic
spin-phonon relaxation times of both simple and complex semiconductors.
Even in the absence of spin-spin scattering that could become more important
at higher carrier concentrations, we find a non-trivial and distinct dependence
of all three spin lifetimes $T_1$, $T_2$ and $T_2^\ast$ on the magnetic field strength.
Future first-principles simulations are necessary to address the additional effects
of electron exchange interactions, including spin-spin scattering on the dephasing
and decoherence of electron spins in semiconductors.

\section*{Acknowledgments}

This work is supported by the Computational Chemical Sciences program within the Office of Science at DOE under Grant No. DE-SC0023301. Calculations were carried out at the Center for Computational Innovations at Rensselaer Polytechnic Institute, and the National Energy Research Scientific Computing Center (NERSC), a U.S. Department of Energy Office of Science User Facility operated under Contract No. DEAC02-05CH11231.

 \end{document}